\documentclass[english,submission]{programming}

\usepackage{makecell}
\usepackage{booktabs}
\usepackage{colortbl}
\usepackage{xcolor}
\usepackage{tikz}
\usepackage{rotating}
\colorlet{higreen}{green!22}
\colorlet{hiyellow}{yellow!38}
\colorlet{hired}{red!16}
\newcommand{\spk}[4]{%
 \begin{tikzpicture}[x=4pt,y=1.2pt,baseline=3pt,inner sep=0pt]
   \fill[black!8](0,0)rectangle(3,10);
   \draw[blue!65!black,line width=0.7pt](0,#1)--(1,#2)--(2,#3)--(3,#4);
   \filldraw[blue!65!black](3,#4)circle(0.9pt);
 \end{tikzpicture}}

\usepackage{listings}
\lstdefinelanguage{Shplait}{%
  keywords=[1]{fun,def,let,type,match,if,cond,block,error,else},%
  keywords=[3]{Int,String,Boolean,Symbol,Listof,Mapof,Vectorof},%
  sensitive=true,%
  morecomment=[l]{//},%
  morestring=[b]",%
  literate=%
    {~typed}{{\textit{\textasciitilde typed}}}{7}%
    {~untyped}{{\textit{\textasciitilde untyped}}}{9}%
    {~else}{{\textit{\textasciitilde else}}}{6}%
    {~type\_source\_mode}{%
      {\textit{\textasciitilde type\_source\_mode}}}{18},%
}
\usepackage{etoolbox}
\AtEndPreamble{\usepackage[capitalise,nameinlink]{cleveref}}

\newcounter{rq}
\newcommand{\RQ}[1]{%
  \refstepcounter{rq}%
  \medskip
  \noindent\textbf{RQ\therq.}\enspace\textit{#1}%
  \par\medskip
}
\newcommand{\RQref}[1]{RQ~\ref{rq:#1}}
\newcommand{\qwen}{qwen2.5-coder:14b}
\newcommand{\haiku}{claude-haiku-4.5}

\paperdetails{
  perspective=engineering,
  area={Types, Debugging, Agents, AI Coding},
}

\begin{document}

\title{Type-Error Ablation and AI Coding Agents}

\author{Shriram Krishnamurthi}[0000-0001-5184-1975]
\affiliation{Brown University, USA}
\author{Matthew Flatt}[0000-0001-8994-2552]
\affiliation{University of Utah, USA}



\maketitle

\begin{abstract}

\paragraph{Context}
For decades, programming language implementors have designed
error messages with one consumer in mind: the human programmer.
Human-factors research has consistently found that programmers
engage with error messages poorly---they skim, miss key
information, and are easily overwhelmed. The practical
consequence has been a strong design pressure toward brevity:
messages should be terse enough that programmers will actually
read them.

\paragraph{Inquiry}
AI coding agents are now a second, fundamentally different
consumer of error messages. Unlike humans, agents do not tire,
lose attention, or find length cognitively overwhelming. The
only practical limit on message size is the context window,
and modern context windows are large. This raises a question
the programming-language community has not previously had
reason to ask: should error-message detail be calibrated
differently for AI agents than for humans?

\paragraph{Approach}
We investigate this question through a controlled experiment
using Shplait, an ML-style statically typed language. We
construct a suite of programs containing a single deliberate
type error each, and measure how often an AI agent repairs them
under ablation: a detailed
error context using the unification stack; a proximate error location;
a minimal type error; and a dynamic (test suite) error only.
An automated oracle uses a test suite to classify each repair attempt
as a type error, semantically incorrect, or semantically correct.

\paragraph{Knowledge}
We find concrete evidence that more detailed error
messages generally improve an agent's ability to fix type errors. We also
find that the presence of a type system appears to help more than only
test suite failure reports.
As a secondary finding, in
cases where an agent successfully fixes the type error, the
resulting program passes all semantic tests most of the
time---lending empirical support to a widely held folk belief
about typed languages. We also see evidence that leading agents are
able to correctly reconstruct the meaning of programs in which all
names have been obfuscated.

\paragraph{Grounding}
Our primary experiment comprises 2,400 trials across ten complete runs, using
{\qwen} via ollama and aider on dedicated GPU hardware. In addition,
we make two more full rounds using {\haiku}. The experimental
platform---correct programs, their erroneous counterparts,
error modes, and oracle---is
described in full.

\paragraph{Importance}
If more detailed error messages genuinely help AI agents, then
languages designed with only human readers in mind may be
providing a suboptimal experience for AI coding assistants.
Our results suggest it is worth considering offering two reporting
modes: a terse human mode and a detailed AI mode. More broadly,
we argue that language-model-targeted error-message design is an open
and important problem for the programming-language community.

\end{abstract}

\section{Introduction: Is More Better?}
\label{sec:intro}

Programming language implementors have long wrestled with how
to present error information to programmers. A type error, for
instance, can arise from a chain of inference steps spanning
many expressions, and presenting the full chain would produce a
wall of text. Decades of human-factors research on error
messages (\cref{sec:rel-work}) have shown that programmers engage with
them surprisingly poorly: they often ignore them, skim past the
key line, or give up when confronted with too much information.
The practical lesson implementors have drawn is that error
messages should be short, pointed, and scannable. The design
space has therefore been shaped almost entirely by the
limitations of the human reader.

That assumption is now worth revisiting. AI coding
agents---tools that read compiler output and iteratively revise
code to fix it---are rapidly becoming a standard part of the
software development workflow. An agent has a fundamentally
different relationship with text than a human does: it does not
get bored or skim, and it is not overwhelmed by
length. In principle, at least, an agent can use all the information
in an error message. The only hard limit is
the context window, and modern context windows can comfortably fit
much or all of the information generated when finding an error.

This observation raises a natural question: \emph{Does the detail
  of an error message affect how well an AI agent can fix the error?}
If it does, then languages designed with only human readers in mind
may be missing out on opportunities.  Perhaps languages should offer
two error-reporting modes---a terse human mode, tuned for the
cognitive limits of programmers, and a more detailed AI mode, tuned for
agents.

It is not, however, obvious that ``more is better''. More detail could
lead an agent into the wrong part of the program rather than toward
the actual mistake. Also, models trained on terse, human-readable
compiler output may not even know what to make of lengthy
diagnostics. In short, this is a question we need to investigate
empirically.

However, it may not be obvious how to set up this study using a
textbook, syntax-directed type-checker. First, the reported location
is often a good place to perform a fix. Second, even if it were not,
it is not immediately clear how to generate a more detailed message in
a principled manner.  As a result, an ablation study is either
impossible or at least meaningless.

To address this, we consider a special
case of errors: those from Hindley-Milner-style type inference. Being
unification-based, they have long been known to sometimes produce
confusing output that blames the wrong code
fragment (\cref{sec:rel-work}). That is,
the error report indicates where a problem was
\emph{detected}, which can be far from where the problem was
\emph{introduced}.  We can instead consider a version of it that
presents not just the proximate error location but the entire
unification stack. This sets up an interesting tension: a brief but
potentially misleading message versus a verbose message more likely to
contain the actual error source. This setup is especially interesting
because a growing number of rich type systems have a similar flavor:
e.g., trait-solving in Rust, which is now the subject of its own
error-debugging research~\cite{gck:int-debug-rust-trait-errors}.

The experimental setup for this work is slightly tricky. We need a
language where we can carefully control for the different output
modes. It is also essential to our experimental setup
(\cref{sec:chaffs}) that we be able to run programs without
type-checking them and halt with safety errors, but most typed
languages are not designed to run in this manner (e.g., their runtime
systems depend on types for safety, and therefore miss several safety
checks). While gradually-typed languages do permit ``turning off the
type checker'', few production ones implement a
Hindley-Milner system; they also tend to have type escape hatches like
\lstinline|Any| that complicate the typed-untyped question.

Instead, we conducted our experiments
by modifying Shplait (\cref{sec:shplait}), an ML-style statically typed language
used in programming languages courses to write definitional
interpreters and similar programs. We adapted Shplait to offer four
``typed'' error output modes: the detailed unification stack, the
conventional unification error, a minimal type error, and no type
checking. (In all four cases, once the program passes the
type-checker---if any---it will produce detailed test suite feedback, including
reports on every test failure.)
We constructed a corpus of
programs each containing a single deliberate type error, and measured
how often an AI agent repaired them under these four modes.
An automated oracle judged each repair attempt, classifying
the result as still a type error, semantically incorrect, or fully
correct with respect to the tests. We go into more detail in
\cref{sec:oracle}. Our key question is one about ablating the amount
of type information:

\RQ{To what extent does the detail in type-error messages
    affect an AI agent's ability to fix them?}%
\label{rq:verbosity}

Two further questions arise naturally. In typed conditions an
agent can test each candidate repair against the type checker,
which is quick; in the untyped condition the only
feedback is from testing---which is slower. Furthermore, while the type
covers all possible runs, it does not provide concrete information
about the failure: many changes can make the program
type-correct and yet fail to be semantically correct. This leads us to
the second question, which compares the untyped case against the typed
ones:

\RQ{How does the presence of a type system in the feedback
    loop affect an agent's ability to converge on a correct
    fix?}%
\label{rq:types}

Finally, there is a widely held
folk belief among typed-language practitioners that our setup lets us
examine directly: if it
type-checks, it probably works. We can investigate this question in
the context of agents:

\RQ{When an agent successfully fixes a type error, to what
    extent does the result also pass its semantic tests?}%
\label{rq:folk}

We find positive evidence that more detailed error
messages generally improve agent fix rates. We also find that type information is
more useful than test failure information.
Finally, we also find strikingly strong support for the folk
belief: 97.9\,\% of cases in which an agent fixed the type error
also passed all semantic tests.

These are all preliminary results
in a very limited setting; we do not claim to have settled
any of these questions. What we do claim is that the questions
are worth asking, that our platform presents an initial
direction for an experimental setup, and that the early evidence
points in an interesting direction.

\section{Background}
\label{sec:background}

\subsection{Language Choice}
\label{sec:shplait}

As noted in \cref{sec:intro}, it can be slightly tricky to choose
a good language for this experiment. We need the ability to control
the error reporting mechanism and introduce modes for different levels
of verbosity. We also need the ability to run programs without the
static type system. This is not as simple as just ``turning off the
type-checker''. If the underlying implementation does not have a
reasonable set of safety checks in place, then type-incorrect programs
might just compute nonsense, generate segmentation faults, and so
on. While these would produce a signal to the agent that the
program is not correct, they are not a helpful diagnostic that make
for a fair comparison against a static type error.

Shplait is built atop Racket~\cite{fffkbmt:programmable-prog-lang}, taking advantage of its
\lstinline|#lang| framework. It is based on the Plait language, which
is designed for the book
\emph{Programming Languages: Application and Interpretation}~\cite{sk:plai3e}
(hence the ``plai'' in the language names; the ``t'' stands for
``typed''). Plait is an SML-inspired language with algebraic datatypes
and Hindley-Milner type inference, and is used at several universities
to build interpreters, type-checkers, type inference engines,
rudimentary compilers, and the like. It is also used at some
institutions to introduce students to typed functional
programming. Plait's default (only) error output is
verbose---effectively what is called
\lstinline|all| in this paper (\cref{sec:modes}).
The user experience with this has been
mixed: sometimes the number of expressions highlighted
can be overwhelming, while at other times it is very helpful
to see multiple expressions to identify the error.

In our experience, however---and as formative experiments in an early
version of this study confirmed---LLMs have some difficulty generating
properly parenthesized code. We wanted to avoid heuristics that
try to correct code that does not even pass the reader. The Shplait
language is essentially Plait but in Shrubbery syntax~\cite{flatt2023rhombus}.
Shrubbery is infix and indentation-based, giving Shplait a
fairly traditional feel.
As an illustrative
example, here is a polymorphic datatype declaration:
\begin{lstlisting}
type AVL(?a)
| avl_empty()
| avl_node(key :: Int, val :: ?a, ht :: Int,
           left :: AVL(?a), right :: AVL(?a))
\end{lstlisting}
and a function with no type annotations (i.e., all types are
inferred):
\begin{lstlisting}
fun avl_lookup(k, t):
  match t
  | avl_empty(): none()
  | avl_node(ck, cv, _h, l, r):
      cond
      | k < ck: avl_lookup(k, l)
      | k > ck: avl_lookup(k, r)
      | ~else:  some(cv)
\end{lstlisting}
We have found that language models can successfully
generate Shplait once supplied with its manual, despite there
being little to no Shplait code in the wild.

Shplait inherits Plait's verbose error reporting default. Like Plait,
it also supports running programs without type-checking.  For this
research project, Shplait was extended with the two additional
type-error--reporting modes described in \cref{sec:modes}.

\subsection{AI Coding Agents}
\label{sec:agents}

For this study, we decided to use a free, open-weight, coding-centric
language model. We justify our choice as follows.

Using free, open-weight models gives us the highest level of
reproducibility. We can archive the model weights with the rest of the
artifact. Anyone (with reasonable hardware) can run it without
requiring paid accounts (which may even be problematic to obtain for
people in certain countries). We need not worry about companies
deprecating older models or quietly changing their behavior.

By way of comparison, a preliminary run (1/10 of what we study in this
paper) of an earlier version of this
study cost about USD~50 in Anthropic
credits. These are not costs that can easily be borne by all
researchers. Furthermore, it is unclear how model usage pricing costs
will evolve. In contrast, free, open-weight models will always remain
available and may see greater uptake in the future.

Ultimately, we decided to choose a model that could complete a run
across multiple nights on a personal laptop (MacBook Air M2 2022 with
24~GiB RAM). This enables us to test out the experimental processes,
confirm that the outputs are coherent, and so on. After a few
experiments, we found that {\qwen}~\cite{hui2024qwen} works
well for our task. It is code-specialized, runs locally via
ollama~\cite{ollama}
without a network dependency, and fits comfortably within the
VRAM budget of our cluster nodes. It is worth noting
(\cref{sec:results}) that this model was, in many cases, able to
successfully fix Shplait code in a single turn.

Since we wanted to simulate a coding \emph{agent},
we use aider~\cite{aider} as the scaffolding layer. Aider is an open-source
AI pair-programming tool that manages the interaction loop between the
LLM and the file system. It presents the prompt to the model, receives
the model's edit expressed as a SEARCH/REPLACE diff (a format that
specifies the exact text to locate and what to substitute), applies
the edit to \lstinline|code.rhm|, runs the oracle, and feeds the
verdict back as the model's next input. We chose aider rather than
writing our own loop to get a well-tested scaffolding layer with a
deterministic edit protocol.

\section{Experiment Design}
\label{sec:design}

\subsection{Programs}
\label{sec:programs}

Our corpus consists of ten programs drawn from standard undergraduate
coursework in data structures, algorithms, and programming
languages. Each is a correct, fully tested Shplait implementation---a
\emph{wheat}~\cite{wk:examplar}---from which the broken variants
(\cref{sec:chaffs}) are derived.  The programs span four thematic
areas: classic data structures (AVL tree, functional queue, leftist
heap), graph and compression algorithms (Dijkstra, Huffman coding,
topological sort), language-processing tools (an interpreter, a
recursive-descent parser, and a regular-expression matcher), and
mathematical computation (polynomial arithmetic). \Cref{tab:programs}
summarizes each program.

\begin{table}
\caption{The ten wheat programs. Also shown are the lengths of the
  programs and how many functions each one contains.}
\label{tab:programs}
\centering
\begin{tabular}{lrrl}
\hline
Program & LOC & Funs & Description \\
\hline
\lstinline|avl-tree|         & 113 & 12 &
  Balanced binary search trees \\
\lstinline|dijkstra|         &  70 &  7 &
  Shortest paths on weighted directed graphs \\
\lstinline|evaluator|        & 113 &  6 &
  Interpreter for a small cbv functional language \\
\lstinline|functional-queue| &  58 &  8 &
  Purely functional queue via the two-list trick \\
\lstinline|huffman|          &  98 & 11 &
  Optimal prefix-free compression \\
\lstinline|leftist-heap|     &  71 &  9 &
  Min-priority queue (leftist heap) \\
\lstinline|poly-arith|       &  66 &  9 &
  Polynomial arithmetic over coefficient lists \\
\lstinline|rd-parser|        & 106 &  7 &
  Recursive-descent parser for arithmetic exprs. \\
\lstinline|regex|            &  54 &  3 &
  Regular-expression matching \\
\lstinline|topo-sort|        &  73 &  6 &
  DFS topological sort with cycle detection \\
\hline
\end{tabular}
\end{table}

In the programs, the datatypes have type annotations (polymorphic for
the generic container types). In contrast,
the programs intentionally carry no type annotations, so as to force
the full use of Hindley-Milner inference and create the most potential
for error message confusion. Every wheat program passes the
type-checker. In addition, every wheat is accompanied by a test suite
that it passes in full; the oracle (\cref{sec:oracle}) uses these same
tests to judge agent repairs.

While we attempted to make these programs correct with respect to
their intended behavior (e.g., that \lstinline|dijkstra| really did
implement Dijkstra's algorithm), in some sense it does not
entirely matter. What matters is that the program passes its test
suite. Any program that passes the same test suite would then be
considered correct.\footnote{This is
  not \emph{entirely} true. Language models have some knowledge of
  these problems, so it does matter if the implementation is wildly
  different from the correct behavior, since the model's knowledge and
  tests would contradict each other, confounding the program fixing
  process. Thus, it matters that the programs are close to correct.
  See also \cref{sec:found-mod}.}

\subsection{Incorrect Programs (Chaffs)}
\label{sec:chaffs}

We derive six chaffs from each wheat, for 60 in total.  A \emph{chaff}~\cite{wk:examplar}
is a broken variant obtained by a local edit to the wheat, leaving the
rest of the program intact. As a concrete example, one
\lstinline|avl-tree| chaff replaces
\begin{lstlisting}
avl_height(avl_left(t)) - avl_height(avl_right(t))
\end{lstlisting}
with
\begin{lstlisting}
avl_left(t) - avl_height(avl_right(t))
\end{lstlisting}
by dropping one call to \lstinline|avl_height| so
the subtraction receives an \lstinline|AVL| tree where it
expects an \lstinline|Int|.

Each chaff has the following characteristics:
\begin{enumerate}

\item It must result in a type error. That is, it cannot purely be a
  logic error that just passes through the type system (e.g.,
  replacing \lstinline|+| with \lstinline|-|).

\item It must fail at least five tests in the test suite. The test
  suite was continuously hardened until this was true for every
  chaff. The tests not only give the agent a clear target, having
  several failures helps to avoid ``fluke'' fixes. In particular, this
  means that the error must be in \emph{live} code with respect to the
  tests. Otherwise the agent can replace the type-offending code with
  \emph{any} code that passes the type-checker---e.g., if the type
  error reports that it was expecting a number, the agent could simply
  replace that entire expression with \lstinline|42|---which would
  then pass the test suite, making it look like the program had
  successfully been fixed.\footnote{Though evident in retrospect, we
    unfortunately discovered this by trial-and-error. An earlier
    version of this study produced what appeared to be absurd
    results. When we examined the ``diff''s between the chaffs and the
    ``fixed'' code, we realized that the edits were meaningless but
    had succeeded because they were in dead-to-the-test-suite code.}
  Observe that Shplait's ability to run the program with the
  type-checker off is essential for obtaining this kind of test
  failure coverage.

\item The single-error constraint is deliberate. With more than one
  simultaneous defect, it becomes harder to tell whether the agent is
  making progress or not; this setup leads to a binary decision on
  whether or not the agent's edit has fixed the program. Obviously,
  real programs may have multiple errors. That said, when code is
  being developed with continuous testing against a quality suite, in
  many cases there will likely be only one (recently-introduced)
  error.

\end{enumerate}

The error edits fall into the following types: wrong-type operand (a
subexpression of the wrong type used in an operator), wrong return
value (a branch or match arm evaluating to the wrong type), branch
mismatch (\lstinline|if| or \lstinline|cond| arms with incompatible
result types), function confusion (one function applied where another
is expected), and wrong-type argument (the wrong value but derived
from the right source). \Cref{tab:chaffs-avl} shows all six
\lstinline|avl-tree| chaffs as a representative example.

Some programs call for symmetric chaff pairs: when a type error
involves a binary operation, the type-checker may assign blame to
either operand depending on its traversal order. Depending on which
branch is blamed, the agent might have an easier or harder time fixing
the bug. We therefore create a pair of chaffs so that if there is an
asymmetry, the more difficult case is guaranteed to be in the
suite. These are the tests labeled as ``mirrored'' in the table.

\begin{table}
\caption{The six \texttt{avl-tree} chaffs, representative
  of the full taxonomy.  Chaffs~1--2 and~4--5 are
  symmetric pairs.}
\label{tab:chaffs-avl}
\centering
\begin{tabular}{clll}
\hline
\# & Error type & Mutation \\
\hline
1 & Wrong-type operand  &
  \lstinline|avl_left(t)| (\lstinline|AVL|) as left operand of
  \lstinline|-| \\
2 & Wrong-type operand  &
  \lstinline|avl_right(t)| (\lstinline|AVL|) as right operand
  (mirror of~1) \\
3 & Function confusion  &
  \lstinline|cons| instead of \lstinline|append| in
  \lstinline|avl_inorder| \\
4 & Wrong return value  &
  Returns \lstinline|l| (\lstinline|AVL|) instead of \lstinline|h|
  (\lstinline|Int|) \\
5 & Wrong return value  &
  Returns \lstinline|r| (\lstinline|AVL|) instead of \lstinline|h|
  (mirror of~4) \\
6 & Wrong-type argument &
  \lstinline|avl_height(ll)| (\lstinline|Int|) where \lstinline|ll|
  (\lstinline|AVL|) expected \\
\hline
\end{tabular}
\end{table}

\subsection{Reporting Modes}
\label{sec:modes}

We study four modes of error-reporting.  Shplait exposes the three
typed modes through the \lstinline|~type_source_mode| pragma.  We
illustrate the difference with the third chaff of
\lstinline|avl-tree|. In all cases, the agent is obtaining information
from the oracle.

\begin{description}

\item[min] The agent receives the two conflicting types. Our goal was to have
  this mode report only the type error with no error locations, but due to a bug
  that we discovered while re-checking our data analysis, this mode prints zero
  error locations (when unifying with an unresolved type variable) or one error
  location (when unifying with an expression). We call these modes min-no-loc
  and min-with-loc, respectively.

  \begin{lstlisting}[language={}]
  TYPE ERROR:
  typecheck failed: Int vs. Listof(Int)\end{lstlisting}

\item[proximate] The agent receives a single
  expression at which the type conflict was detected, and its source location.
  (On one hand,
  this is no more informative than min-with-loc, since a human can reconstruct
  the expression from the location. But language models are notoriously poor at
  counting, and in addition may not know how to identify the end of the
  expression.)
  Thus, this mode's output is very
  similar to that of
  most ML-family type systems, including Standard~ML and OCaml. The
  exact details of the algorithm are not too interesting; it is
  essentially that described in PLAI~\cite{sk:plai3e} and similar to
  baseline approaches described by Duggan and Bent~\cite{DugganBent1996}.

  \begin{lstlisting}[language={}]
  TYPE ERROR:
  typecheck failed: Int vs. Listof(Int)
    sources:
      cons
    location...:
     code.rhm:87:6\end{lstlisting}

   In 39 of our 60 chaffs, the reported line(s) in
   \lstinline|proximate| mode matched that where the error was
   injected. In the other 21 it did not, with line-number distances
   ranging from 1 (in 5 cases), 2--8 (in 5 more), to 13--25 (in 5
   more), to as high as 44.

 \item[all] The agent receives the entire remaining unification stack,
   along with locations. This can be substantially longer. However, it
   is also more likely to contain the expression that truly introduced
   the error. Each location in the bottom part of the output is that
   of the corresponding source expression in the top part of the
   output (in the listing below, those on lines 4, 5, 6, 7,
   and the compound \lstinline|match| expression starting on 8).

  \begin{lstlisting}[language={}]
  TYPE ERROR:
typecheck failed: Int vs. Listof(Int)
  sources:
    k
    Int
    cons
    match t
| avl_empty(): []
| avl_node(k, _v, _h, l, r):
    cons(avl_inorder(l), cons(k, avl_inorder(r)))
[4 more expressions, 9 lines long]
  location...:
   code.rhm:86:13
   code.rhm:6:18
   code.rhm:87:6
   code.rhm:84:2
   code.rhm:87:32
[4 more lines]\end{lstlisting}

\item[untyped] mode disables the type checker entirely, via Shplait's
  \lstinline|~untyped| pragma. The agent's only feedback is whether
  the test suite passes or fails. This mode isolates a distinct
  question---does having a type system in the repair loop matter at
  all?---from the verbosity question proper. There could be
  multiple test failures; below we show two of the 8 failing tests.

  \begin{lstlisting}[language={}]
  TESTS FAILED:
  avl-tree-tests.rhm:264: failed
    got:      [[], 5]
    expected: [5]
  avl-tree-tests.rhm:271: failed
    got:      [[[], 1], 2, [[], 3], 4, [], 5]
    expected: [1, 2, 3, 4, 5]
  ...
  8/70 test failures\end{lstlisting}

\end{description}

One question is whether the presence of these pragmas might itself
affect the performance of the LLM. To avoid turning this into a
confound, we instead did the following. The second line, where the
pragma goes, is left blank in the code that, along with error output,
is sent to the LLM. After editing the source based on its output, the
pragma is then \emph{injected} into file, and this version is sent to
the oracle. The presence of the blank line means that all the line and
column numbers from the error outputs are the same as those of the
program that the LLM sees, in case they affect the LLM.

\subsection{Oracle}
\label{sec:oracle}

The oracle is a shell script that serves as the test command
for every trial. It takes the current verbosity mode as an
argument and runs from the program's working directory, where
\lstinline|code.rhm| holds the agent's current attempt.

For the three typed modes, the oracle injects the appropriate
\lstinline|~type_source_mode| pragma into line~2 of a
temporary copy of \lstinline|code.rhm| and runs
\lstinline|raco test| on it. A non-zero exit signals a type
error; the oracle returns \textsc{type error} followed by the
diagnostic, stripping Shplait's internal unification-context
stack to remove implementation noise. If the code
type-checks, the oracle runs \lstinline|raco test| on the
program's test file. Failure returns \textsc{tests failed}
with the test output; success returns \textsc{success}. For
the untyped mode, the type-checking step is skipped and
\lstinline|~untyped| is injected instead.

The oracle is deterministic and reproducible: given the same
\lstinline|code.rhm| and mode, it always returns the same
verdict. It judges only the final code, not the agent's
reasoning or intermediate steps. The test suite used for
judging is the same one each wheat passes in full; there is
no separate evaluation set.

\subsection{Agent Setup}
\label{sec:agent-setup}

\begin{figure}
\begin{lstlisting}[language={},basicstyle=\ttfamily\footnotesize]
The file `code.rhm` is a Shplait program that contains a type
error. Your task is to fix it so that:
 1. The program passes the Shplait type checker (no type errors).
 2. The program passes the test suite.

Use the test command to check your work. It will tell you one of:
 - TYPE ERROR: <message>    the type checker rejected the code
 - TESTS FAILED: <output>   type-checks but some tests fail
 - SUCCESS                  both checks pass; you are done

Make the MINIMAL change needed to fix the type error. Do not add
type annotations, refactor, or modify functions not directly
involved in the error.

Shplait uses Hindley-Milner type inference (same as SML/OCaml).
The error message points to where the conflict was detected, which
may not be where the bug is. If the flagged expression looks
correct, look at the other expressions whose types flow into it.

IMPORTANT: Line 2 of `code.rhm` must remain blank. It is reserved
for internal use by the harness. Do not place any code, comments,
or directives there.
\end{lstlisting}
\caption{The prompt given to the agent at the start of every trial.
 The SEARCH/REPLACE edit format is part of aider's protocol and
 known to the model from training; it is not explained in the
 prompt.}
\label{fig:prompt}
\end{figure}

Aider is given
\lstinline|code.rhm| to edit along with two read-only reference files:
the Shplait language documentation and a list of built-in
functions.
It presents the initial prompt
(\cref{fig:prompt}\footnote{The prompt is oriented towards typed
  programs and type errors. We nevertheless kept the prompt constant
  across all runs lest that become a confound, since even small
  changes could have big impacts that would be hard to assess.}) to
the model, applies the model's SEARCH/REPLACE edits to
\lstinline|code.rhm|, invokes the oracle after each edit, and feeds
the verdict back as the next input. The prompt instructs the agent to make a minimal fix,
explains the oracle's three-way output format, and notes that
Hindley-Milner blame may not point to the true error
source~(\cref{sec:modes}). The LLM is not told which verbosity mode
is active.

\subsection{Procedure}
\label{sec:procedure}

The experiment is organized as a series of independent
\emph{runs}. Each run is a full sweep of all
$10 \times 6 \times 4 = 240$ \emph{trials}, one for every
combination of program, chaff, and mode. We report results
from 10 complete runs, for a total of
2400 trials.

A single trial proceeds as follows. The harness copies the
chaff to \lstinline|code.rhm|, replacing line~2 (which carries
the chaff's ~untyped marker so the chaff is runnable standalone)
with a blank line, so the
agent does not see which mode is active. Aider is launched with
the initial prompt and the oracle configured as its test
command. The agent iteratively proposes edits; aider applies
each one, calls the oracle, and returns the verdict to the
agent. The trial ends when the oracle returns
\textsc{success}, the five-minute timeout fires, or aider
gives up. The final state, number of editing turns, and token
counts are written to a JSON file.

Trial independence is enforced at several levels. Within a run, the
harness overwrites \lstinline|code.rhm| with a fresh copy of the chaff
at the start of each trial and deletes the Racket bytecode cache
(\lstinline|compiled/|) before every oracle call, so no compiled state
or edited code carries over between trials. Each trial launches a
fresh aider process with an empty conversation context; the ollama
server is persistent across trials (to avoid reload overhead) but the
model holds no memory between requests, so one trial's conversation
cannot influence the next. The oracle uses a per-invocation temporary
file for the mode-injected code, cleaned up on exit, so concurrent
oracle calls within a node cannot interfere with each other.

The setup makes sure that the actual tests cannot leak into the model
as follows. Only the code file, along with the oracle's output, are
given to aider. Aider is run with \lstinline|--no-git| so as to not
inadvertently pull in the test files through a repository.

To allow runs to execute in parallel on a cluster, each
job operates on a private copy of the \lstinline|programs/|
directory. This is the critical guard against one run clobbering
another's \lstinline|code.rhm| mid-trial. Result files are written to
per-run, per-mode subdirectories, so there is no possibility of one
run overwriting another's output.

\section{Results}
\label{sec:results}

\subsection{Experiment Run}
\label{sec:exp-run}

We present the results of our primary experimental analysis. They are
based on 10 experimental runs, for a total of 2400 trials, using
{\qwen}.

We impose a 600s wall-clock timeout per trial. Aider
additionally stops after three successive test-fail-retry
cycles (its hardcoded limit) or when the model replies without
a code edit. Experiments ran on a Hydra HPC cluster
using NVIDIA data-center GPUs (L40S, L40, A6000, and
TITAN~RTX). We used the Slurm~\cite{yoo2003slurm} cluster manager
to facilitate running jobs.
Nodes with only~11~GiB of VRAM (RTX~2080~Ti)
caused model weights to spill to CPU, introducing a roughly
$2.4\times$ slowdown but also---much more importantly---a systematic
timeout bias; those trials were cancelled and resubmitted to adequate
hardware.

\subsection{\RQref{verbosity} : The Impact of Error Verbosity}
\label{sec:rq-verb}

\begin{figure}[t]
\centering
\includegraphics[width=\linewidth]{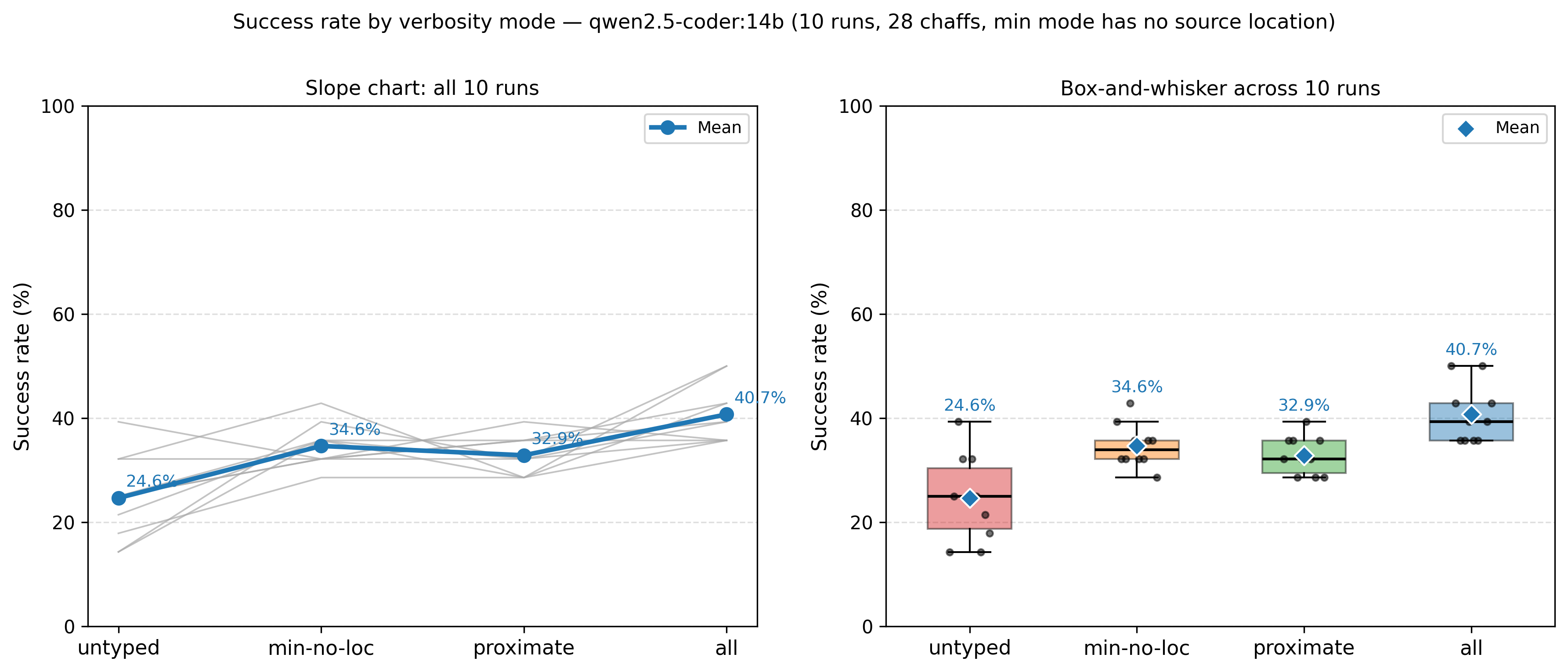}\\[6pt]
\includegraphics[width=\linewidth]{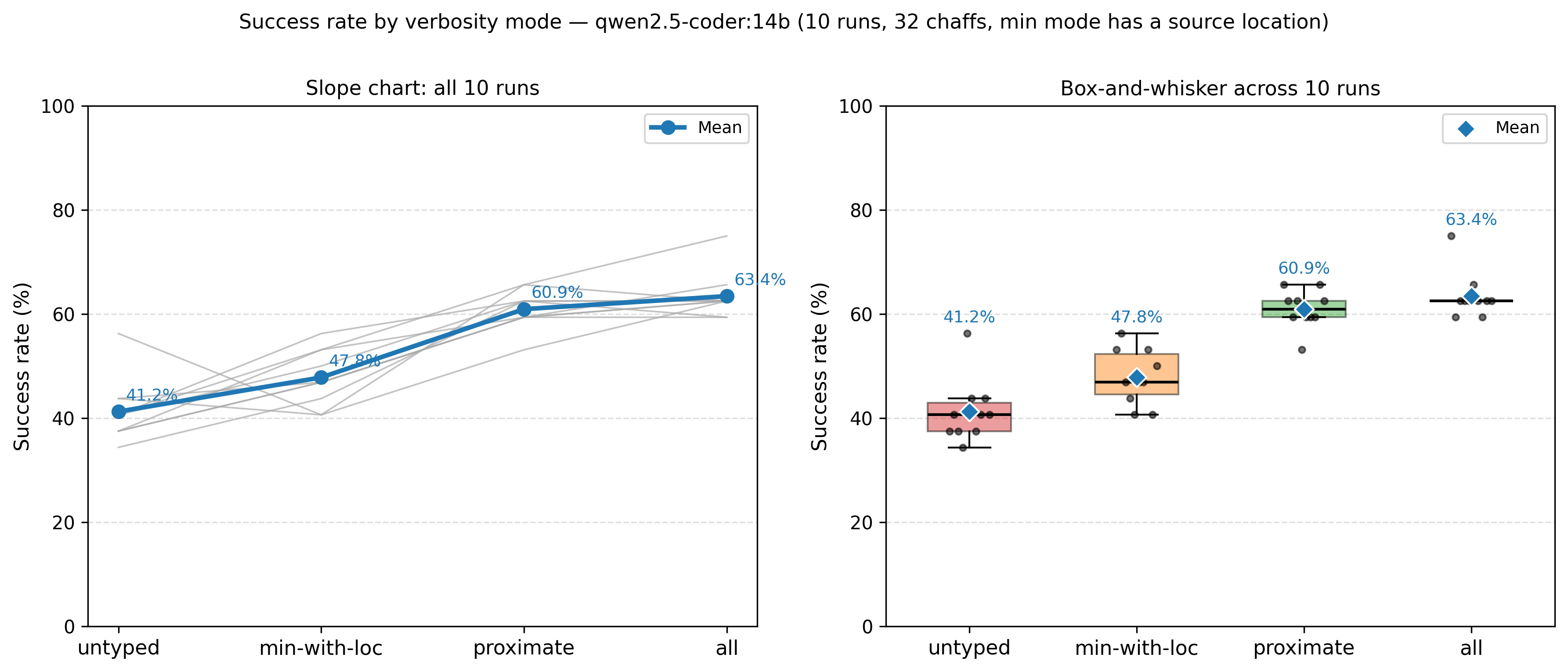}
\caption{
  Top: min-no-loc; bottom: min-with-loc. Left: slope chart showing success
  counts across the four modes (gray = individual runs, blue = mean).
  Right: corresponding box-and-whisker, with individual run rates as dots.}
\label{fig:results}
\end{figure}

\RQref{verbosity} asks whether verbosity level affects fix
rates. \Cref{fig:results} and \cref{tab:outcomes-qwen} summarize the results.
The four modes generally show an upward trend
in mean success rate from \lstinline|untyped| to
\lstinline|min| to \lstinline|proximate| to \lstinline|all|.

We do see differences between the two \lstinline|min| modes. The
\lstinline|min-no-loc| mode corresponds (\cref{sec:modes}) to more distant
errors. This may explain why we see the notable jump from it and
\lstinline|proximate| to \lstinline|all|. In contrast, \lstinline|min-with-loc|
reflects a more proximate error report. This may explain why the notable jump
here is to \lstinline|proximate|, and \lstinline|all| adds little more. Of
course, there is much more noise in individual runs.

When successful, agents rarely needed more than one round of
interaction: the median turns-to-success is 1 across all modes, and
the mean never exceeds 1.3. Success was almost always reached on the
first edit, or not at all. This suggests that the primary
effect of verbosity is not in enabling multi-step reasoning but in
helping the agent form a correct first hypothesis.

\begin{table}[t]
  \centering
  \caption{Outcome distribution for {\qwen}. Each cell shows
    \emph{a}\,/\,\emph{b}, where \emph{a} is the percentage (\%) of that
    mode's trials on the 28 min-no-loc chaffs and \emph{b} is the
    percentage on the 32 min-with-loc chaffs. \emph{Timed out} = the wall clock fired while
    the agent was still working (\emph{no change}: it never applied an
    edit before time ran out). \emph{Halted} = aider's loop ended on its
    own before the clock---\emph{no edit}, the model never produced an
    applicable edit; \emph{stopped}, it applied one or more edits and
    then returned a reply with no further edit, leaving aider nothing to
    apply; \emph{reflection cap}, it reached aider's hard limit of three
    reflection retries.}
  \label{tab:outcomes-qwen}
  \setlength{\tabcolsep}{4pt}
  \begin{tabular}{@{}lr rrr rrr@{}}
    \toprule
    & & \multicolumn{3}{c}{Timed out} & \multicolumn{3}{c}{Halted} \\
    \cmidrule(lr){3-5}\cmidrule(lr){6-8}
    Mode & Success & \makecell{tests\\failed} & \makecell{type\\error} & \makecell{no\\change} & \makecell{no\\edit} & stopped & \makecell{reflection\\cap} \\
    \midrule
    \lstinline|untyped| & 24.6/41.2 & 10.0/0.9 &  \multicolumn{1}{c}{---} & 2.9/1.6 & 13.9/9.7 & 15.7/7.5 & 32.9/39.1 \\
    \lstinline|min-...| & 34.6/47.8 & 0.4/1.6 & 3.2/0.9 & 0.0/0.0 & 1.4/1.9 & 3.6/1.9 & 56.8/45.9 \\
    \lstinline|proximate| & 32.9/60.9 & 1.4/0.6 & 2.1/2.5 & 0.0/0.0 & 8.6/3.4 & 1.1/0.9 & 53.9/31.6 \\
    \lstinline|all| & 40.7/63.4 & 1.1/0.9 & 0.4/1.2 & 0.4/0.0 & 10.4/3.4 & 1.1/1.9 & 46.1/29.1 \\
    \bottomrule
  \end{tabular}
\end{table}

\paragraph{Statistical Analysis}

Given the many limitations of this study, we do not believe there is
value to doing an extensive statistical analysis of these
results. Nevertheless, it is useful to perform a basic test to check
that our initial hypothesis actually holds in these data, as suggested
by the trend line.

We tested the predicted ordering \lstinline|untyped| $<$
\lstinline|min| $<$ \lstinline|proximate| $<$ \lstinline|all|
for each \lstinline|min| mode
using
Page's trend test~\cite{page1963ordered}, which handles repeated measures for
ordered alternatives.  The experimental unit is a (chaff, run) pair;
each such pair yields a binary outcome (success or
failure) under all four modes.  We run the test on each chaff bin
separately, giving a $280 \times 4$ block matrix for min-no-loc
(28 chaffs $\times$ 10 runs) and a $320 \times 4$ block matrix for
min-with-loc (32 chaffs $\times$ 10 runs).\footnote{We treat each run distinctly because the runs took
  place on varying hardware, which could impact factors like
  timeouts. Thus it is most meaningful to treat each of these pairs
  distinctly. All four modes for that pair ran on the same hardware.}
Both tests were highly significant: for min-no-loc, $L = 7130$,
$z = 2.69$, $p = 3.6 \times 10^{-3}$, one-sided; for min-with-loc,
$L = 8255$, $z = 4.94$, $p = 4.0 \times 10^{-7}$, one-sided.
In short, this confirms that the trend we visually observe
is statistically robust in both bins.

\paragraph{Variance Across Chaffs}


\begin{figure}[p]
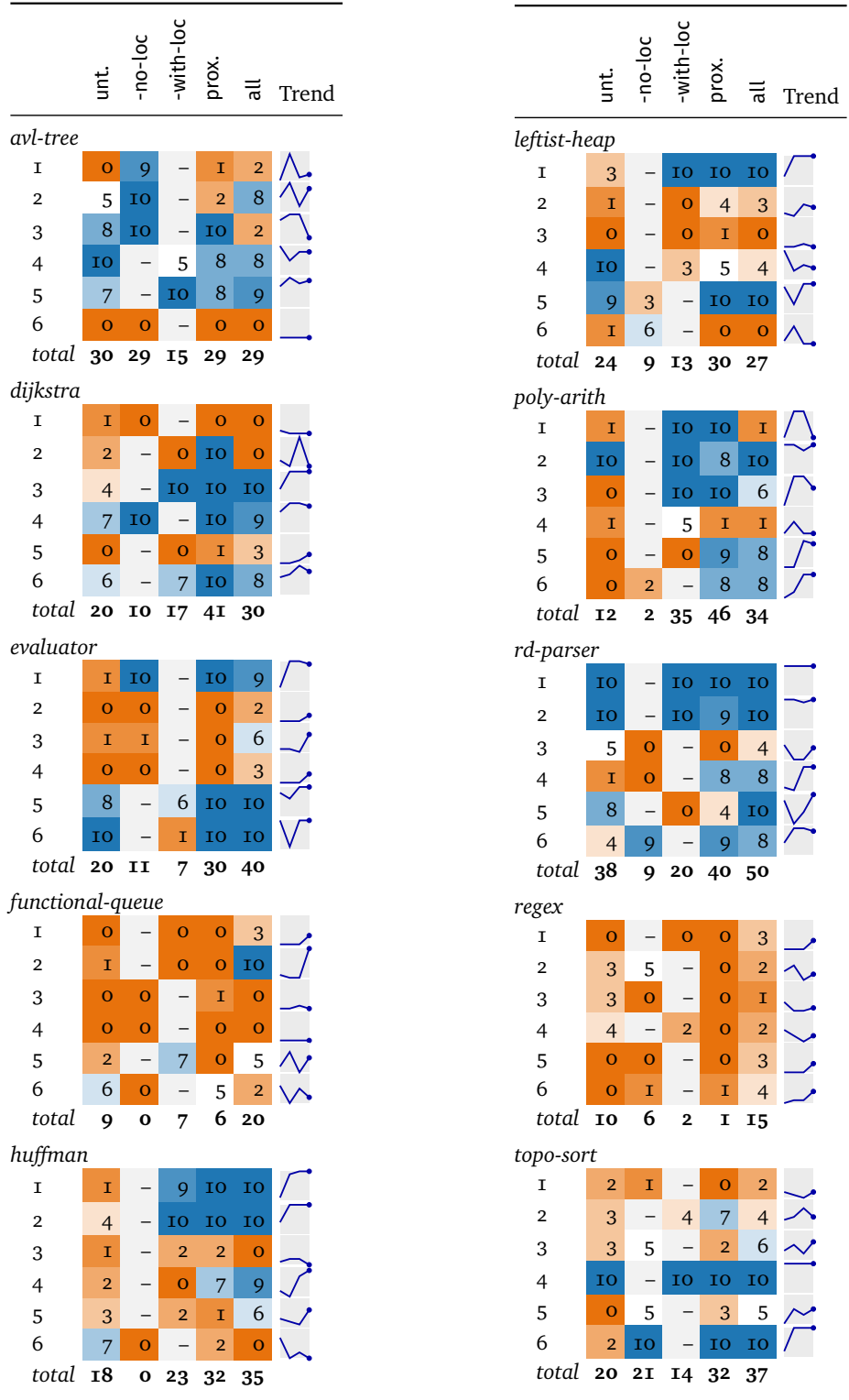

\centering
\footnotesize
\setlength{\tabcolsep}{3pt}
\renewcommand{\arraystretch}{1.05}

\begin{minipage}[t]{0.48\linewidth}
\centering
\begin{tabular}{@{}l rrrrr l@{}}
\toprule
   & \multicolumn{1}{c}{\rotatebox{90}{\texttt{unt.}\;}} & \multicolumn{1}{c}{\rotatebox{90}{\texttt{-no-loc}\;}} & \multicolumn{1}{c}{\rotatebox{90}{\texttt{-with-loc}\;}} & \multicolumn{1}{c}{\rotatebox{90}{\texttt{prox.}\;}} & \multicolumn{1}{c}{\rotatebox{90}{\texttt{all}\;}} & \multicolumn{1}{c}{Trend}
\\ \midrule
\multicolumn{7}{@{}l}{\textit{avl-tree}} \\[-1pt]
  \quad 1 & \cellcolor[HTML]{E8720C}0 & \cellcolor[HTML]{4B92C3}9 & \cellcolor[HTML]{F2F2F2}{\scriptsize--} & \cellcolor[HTML]{EC8E3C}1 & \cellcolor[HTML]{F1AA6D}2 & \spk{0}{9}{1}{2} \\
  \quad 2 & \cellcolor[HTML]{FFFFFF}5 & \cellcolor[HTML]{1F77B4}10 & \cellcolor[HTML]{F2F2F2}{\scriptsize--} & \cellcolor[HTML]{F1AA6D}2 & \cellcolor[HTML]{78ADD2}8 & \spk{5}{10}{2}{8} \\
  \quad 3 & \cellcolor[HTML]{78ADD2}8 & \cellcolor[HTML]{1F77B4}10 & \cellcolor[HTML]{F2F2F2}{\scriptsize--} & \cellcolor[HTML]{1F77B4}10 & \cellcolor[HTML]{F1AA6D}2 & \spk{8}{10}{10}{2} \\
  \quad 4 & \cellcolor[HTML]{1F77B4}10 & \cellcolor[HTML]{F2F2F2}{\scriptsize--} & \cellcolor[HTML]{FFFFFF}5 & \cellcolor[HTML]{78ADD2}8 & \cellcolor[HTML]{78ADD2}8 & \spk{10}{5}{8}{8} \\
  \quad 5 & \cellcolor[HTML]{A5C8E1}7 & \cellcolor[HTML]{F2F2F2}{\scriptsize--} & \cellcolor[HTML]{1F77B4}10 & \cellcolor[HTML]{78ADD2}8 & \cellcolor[HTML]{4B92C3}9 & \spk{7}{10}{8}{9} \\
  \quad 6 & \cellcolor[HTML]{E8720C}0 & \cellcolor[HTML]{E8720C}0 & \cellcolor[HTML]{F2F2F2}{\scriptsize--} & \cellcolor[HTML]{E8720C}0 & \cellcolor[HTML]{E8720C}0 & \spk{0}{0}{0}{0} \\
  \quad\textit{total} & \textbf{30} & \textbf{29} & \textbf{15} & \textbf{29} & \textbf{29} & \\[3pt]
\multicolumn{7}{@{}l}{\textit{dijkstra}} \\[-1pt]
  \quad 1 & \cellcolor[HTML]{EC8E3C}1 & \cellcolor[HTML]{E8720C}0 & \cellcolor[HTML]{F2F2F2}{\scriptsize--} & \cellcolor[HTML]{E8720C}0 & \cellcolor[HTML]{E8720C}0 & \spk{1}{0}{0}{0} \\
  \quad 2 & \cellcolor[HTML]{F1AA6D}2 & \cellcolor[HTML]{F2F2F2}{\scriptsize--} & \cellcolor[HTML]{E8720C}0 & \cellcolor[HTML]{1F77B4}10 & \cellcolor[HTML]{E8720C}0 & \spk{2}{0}{10}{0} \\
  \quad 3 & \cellcolor[HTML]{FAE2CE}4 & \cellcolor[HTML]{F2F2F2}{\scriptsize--} & \cellcolor[HTML]{1F77B4}10 & \cellcolor[HTML]{1F77B4}10 & \cellcolor[HTML]{1F77B4}10 & \spk{4}{10}{10}{10} \\
  \quad 4 & \cellcolor[HTML]{A5C8E1}7 & \cellcolor[HTML]{1F77B4}10 & \cellcolor[HTML]{F2F2F2}{\scriptsize--} & \cellcolor[HTML]{1F77B4}10 & \cellcolor[HTML]{4B92C3}9 & \spk{7}{10}{10}{9} \\
  \quad 5 & \cellcolor[HTML]{E8720C}0 & \cellcolor[HTML]{F2F2F2}{\scriptsize--} & \cellcolor[HTML]{E8720C}0 & \cellcolor[HTML]{EC8E3C}1 & \cellcolor[HTML]{F5C69D}3 & \spk{0}{0}{1}{3} \\
  \quad 6 & \cellcolor[HTML]{D2E3F0}6 & \cellcolor[HTML]{F2F2F2}{\scriptsize--} & \cellcolor[HTML]{A5C8E1}7 & \cellcolor[HTML]{1F77B4}10 & \cellcolor[HTML]{78ADD2}8 & \spk{6}{7}{10}{8} \\
  \quad\textit{total} & \textbf{20} & \textbf{10} & \textbf{17} & \textbf{41} & \textbf{30} & \\[3pt]
\multicolumn{7}{@{}l}{\textit{evaluator}} \\[-1pt]
  \quad 1 & \cellcolor[HTML]{EC8E3C}1 & \cellcolor[HTML]{1F77B4}10 & \cellcolor[HTML]{F2F2F2}{\scriptsize--} & \cellcolor[HTML]{1F77B4}10 & \cellcolor[HTML]{4B92C3}9 & \spk{1}{10}{10}{9} \\
  \quad 2 & \cellcolor[HTML]{E8720C}0 & \cellcolor[HTML]{E8720C}0 & \cellcolor[HTML]{F2F2F2}{\scriptsize--} & \cellcolor[HTML]{E8720C}0 & \cellcolor[HTML]{F1AA6D}2 & \spk{0}{0}{0}{2} \\
  \quad 3 & \cellcolor[HTML]{EC8E3C}1 & \cellcolor[HTML]{EC8E3C}1 & \cellcolor[HTML]{F2F2F2}{\scriptsize--} & \cellcolor[HTML]{E8720C}0 & \cellcolor[HTML]{D2E3F0}6 & \spk{1}{1}{0}{6} \\
  \quad 4 & \cellcolor[HTML]{E8720C}0 & \cellcolor[HTML]{E8720C}0 & \cellcolor[HTML]{F2F2F2}{\scriptsize--} & \cellcolor[HTML]{E8720C}0 & \cellcolor[HTML]{F5C69D}3 & \spk{0}{0}{0}{3} \\
  \quad 5 & \cellcolor[HTML]{78ADD2}8 & \cellcolor[HTML]{F2F2F2}{\scriptsize--} & \cellcolor[HTML]{D2E3F0}6 & \cellcolor[HTML]{1F77B4}10 & \cellcolor[HTML]{1F77B4}10 & \spk{8}{6}{10}{10} \\
  \quad 6 & \cellcolor[HTML]{1F77B4}10 & \cellcolor[HTML]{F2F2F2}{\scriptsize--} & \cellcolor[HTML]{EC8E3C}1 & \cellcolor[HTML]{1F77B4}10 & \cellcolor[HTML]{1F77B4}10 & \spk{10}{1}{10}{10} \\
  \quad\textit{total} & \textbf{20} & \textbf{11} & \textbf{7} & \textbf{30} & \textbf{40} & \\[3pt]
\multicolumn{7}{@{}l}{\textit{functional-queue}} \\[-1pt]
  \quad 1 & \cellcolor[HTML]{E8720C}0 & \cellcolor[HTML]{F2F2F2}{\scriptsize--} & \cellcolor[HTML]{E8720C}0 & \cellcolor[HTML]{E8720C}0 & \cellcolor[HTML]{F5C69D}3 & \spk{0}{0}{0}{3} \\
  \quad 2 & \cellcolor[HTML]{EC8E3C}1 & \cellcolor[HTML]{F2F2F2}{\scriptsize--} & \cellcolor[HTML]{E8720C}0 & \cellcolor[HTML]{E8720C}0 & \cellcolor[HTML]{1F77B4}10 & \spk{1}{0}{0}{10} \\
  \quad 3 & \cellcolor[HTML]{E8720C}0 & \cellcolor[HTML]{E8720C}0 & \cellcolor[HTML]{F2F2F2}{\scriptsize--} & \cellcolor[HTML]{EC8E3C}1 & \cellcolor[HTML]{E8720C}0 & \spk{0}{0}{1}{0} \\
  \quad 4 & \cellcolor[HTML]{E8720C}0 & \cellcolor[HTML]{E8720C}0 & \cellcolor[HTML]{F2F2F2}{\scriptsize--} & \cellcolor[HTML]{E8720C}0 & \cellcolor[HTML]{E8720C}0 & \spk{0}{0}{0}{0} \\
  \quad 5 & \cellcolor[HTML]{F1AA6D}2 & \cellcolor[HTML]{F2F2F2}{\scriptsize--} & \cellcolor[HTML]{A5C8E1}7 & \cellcolor[HTML]{E8720C}0 & \cellcolor[HTML]{FFFFFF}5 & \spk{2}{7}{0}{5} \\
  \quad 6 & \cellcolor[HTML]{D2E3F0}6 & \cellcolor[HTML]{E8720C}0 & \cellcolor[HTML]{F2F2F2}{\scriptsize--} & \cellcolor[HTML]{FFFFFF}5 & \cellcolor[HTML]{F1AA6D}2 & \spk{6}{0}{5}{2} \\
  \quad\textit{total} & \textbf{9} & \textbf{0} & \textbf{7} & \textbf{6} & \textbf{20} & \\[3pt]
\multicolumn{7}{@{}l}{\textit{huffman}} \\[-1pt]
  \quad 1 & \cellcolor[HTML]{EC8E3C}1 & \cellcolor[HTML]{F2F2F2}{\scriptsize--} & \cellcolor[HTML]{4B92C3}9 & \cellcolor[HTML]{1F77B4}10 & \cellcolor[HTML]{1F77B4}10 & \spk{1}{9}{10}{10} \\
  \quad 2 & \cellcolor[HTML]{FAE2CE}4 & \cellcolor[HTML]{F2F2F2}{\scriptsize--} & \cellcolor[HTML]{1F77B4}10 & \cellcolor[HTML]{1F77B4}10 & \cellcolor[HTML]{1F77B4}10 & \spk{4}{10}{10}{10} \\
  \quad 3 & \cellcolor[HTML]{EC8E3C}1 & \cellcolor[HTML]{F2F2F2}{\scriptsize--} & \cellcolor[HTML]{F1AA6D}2 & \cellcolor[HTML]{F1AA6D}2 & \cellcolor[HTML]{E8720C}0 & \spk{1}{2}{2}{0} \\
  \quad 4 & \cellcolor[HTML]{F1AA6D}2 & \cellcolor[HTML]{F2F2F2}{\scriptsize--} & \cellcolor[HTML]{E8720C}0 & \cellcolor[HTML]{A5C8E1}7 & \cellcolor[HTML]{4B92C3}9 & \spk{2}{0}{7}{9} \\
  \quad 5 & \cellcolor[HTML]{F5C69D}3 & \cellcolor[HTML]{F2F2F2}{\scriptsize--} & \cellcolor[HTML]{F1AA6D}2 & \cellcolor[HTML]{EC8E3C}1 & \cellcolor[HTML]{D2E3F0}6 & \spk{3}{2}{1}{6} \\
  \quad 6 & \cellcolor[HTML]{A5C8E1}7 & \cellcolor[HTML]{E8720C}0 & \cellcolor[HTML]{F2F2F2}{\scriptsize--} & \cellcolor[HTML]{F1AA6D}2 & \cellcolor[HTML]{E8720C}0 & \spk{7}{0}{2}{0} \\
  \quad\textit{total} & \textbf{18} & \textbf{0} & \textbf{23} & \textbf{32} & \textbf{35} & \\[3pt]
\bottomrule
\end{tabular}
\end{minipage}%
\hfill
\begin{minipage}[t]{0.48\linewidth}
\centering
\begin{tabular}{@{}l rrrrr l@{}}
\toprule
   & \multicolumn{1}{c}{\rotatebox{90}{\texttt{unt.}\;}} & \multicolumn{1}{c}{\rotatebox{90}{\texttt{-no-loc}\;}} & \multicolumn{1}{c}{\rotatebox{90}{\texttt{-with-loc}\;}} & \multicolumn{1}{c}{\rotatebox{90}{\texttt{prox.}\;}} & \multicolumn{1}{c}{\rotatebox{90}{\texttt{all}\;}} & \multicolumn{1}{c}{Trend}
\\ \midrule
\multicolumn{7}{@{}l}{\textit{leftist-heap}} \\[-1pt]
  \quad 1 & \cellcolor[HTML]{F5C69D}3 & \cellcolor[HTML]{F2F2F2}{\scriptsize--} & \cellcolor[HTML]{1F77B4}10 & \cellcolor[HTML]{1F77B4}10 & \cellcolor[HTML]{1F77B4}10 & \spk{3}{10}{10}{10} \\
  \quad 2 & \cellcolor[HTML]{EC8E3C}1 & \cellcolor[HTML]{F2F2F2}{\scriptsize--} & \cellcolor[HTML]{E8720C}0 & \cellcolor[HTML]{FAE2CE}4 & \cellcolor[HTML]{F5C69D}3 & \spk{1}{0}{4}{3} \\
  \quad 3 & \cellcolor[HTML]{E8720C}0 & \cellcolor[HTML]{F2F2F2}{\scriptsize--} & \cellcolor[HTML]{E8720C}0 & \cellcolor[HTML]{EC8E3C}1 & \cellcolor[HTML]{E8720C}0 & \spk{0}{0}{1}{0} \\
  \quad 4 & \cellcolor[HTML]{1F77B4}10 & \cellcolor[HTML]{F2F2F2}{\scriptsize--} & \cellcolor[HTML]{F5C69D}3 & \cellcolor[HTML]{FFFFFF}5 & \cellcolor[HTML]{FAE2CE}4 & \spk{10}{3}{5}{4} \\
  \quad 5 & \cellcolor[HTML]{4B92C3}9 & \cellcolor[HTML]{F5C69D}3 & \cellcolor[HTML]{F2F2F2}{\scriptsize--} & \cellcolor[HTML]{1F77B4}10 & \cellcolor[HTML]{1F77B4}10 & \spk{9}{3}{10}{10} \\
  \quad 6 & \cellcolor[HTML]{EC8E3C}1 & \cellcolor[HTML]{D2E3F0}6 & \cellcolor[HTML]{F2F2F2}{\scriptsize--} & \cellcolor[HTML]{E8720C}0 & \cellcolor[HTML]{E8720C}0 & \spk{1}{6}{0}{0} \\
  \quad\textit{total} & \textbf{24} & \textbf{9} & \textbf{13} & \textbf{30} & \textbf{27} & \\[3pt]
\multicolumn{7}{@{}l}{\textit{poly-arith}} \\[-1pt]
  \quad 1 & \cellcolor[HTML]{EC8E3C}1 & \cellcolor[HTML]{F2F2F2}{\scriptsize--} & \cellcolor[HTML]{1F77B4}10 & \cellcolor[HTML]{1F77B4}10 & \cellcolor[HTML]{EC8E3C}1 & \spk{1}{10}{10}{1} \\
  \quad 2 & \cellcolor[HTML]{1F77B4}10 & \cellcolor[HTML]{F2F2F2}{\scriptsize--} & \cellcolor[HTML]{1F77B4}10 & \cellcolor[HTML]{78ADD2}8 & \cellcolor[HTML]{1F77B4}10 & \spk{10}{10}{8}{10} \\
  \quad 3 & \cellcolor[HTML]{E8720C}0 & \cellcolor[HTML]{F2F2F2}{\scriptsize--} & \cellcolor[HTML]{1F77B4}10 & \cellcolor[HTML]{1F77B4}10 & \cellcolor[HTML]{D2E3F0}6 & \spk{0}{10}{10}{6} \\
  \quad 4 & \cellcolor[HTML]{EC8E3C}1 & \cellcolor[HTML]{F2F2F2}{\scriptsize--} & \cellcolor[HTML]{FFFFFF}5 & \cellcolor[HTML]{EC8E3C}1 & \cellcolor[HTML]{EC8E3C}1 & \spk{1}{5}{1}{1} \\
  \quad 5 & \cellcolor[HTML]{E8720C}0 & \cellcolor[HTML]{F2F2F2}{\scriptsize--} & \cellcolor[HTML]{E8720C}0 & \cellcolor[HTML]{4B92C3}9 & \cellcolor[HTML]{78ADD2}8 & \spk{0}{0}{9}{8} \\
  \quad 6 & \cellcolor[HTML]{E8720C}0 & \cellcolor[HTML]{F1AA6D}2 & \cellcolor[HTML]{F2F2F2}{\scriptsize--} & \cellcolor[HTML]{78ADD2}8 & \cellcolor[HTML]{78ADD2}8 & \spk{0}{2}{8}{8} \\
  \quad\textit{total} & \textbf{12} & \textbf{2} & \textbf{35} & \textbf{46} & \textbf{34} & \\[3pt]
\multicolumn{7}{@{}l}{\textit{rd-parser}} \\[-1pt]
  \quad 1 & \cellcolor[HTML]{1F77B4}10 & \cellcolor[HTML]{F2F2F2}{\scriptsize--} & \cellcolor[HTML]{1F77B4}10 & \cellcolor[HTML]{1F77B4}10 & \cellcolor[HTML]{1F77B4}10 & \spk{10}{10}{10}{10} \\
  \quad 2 & \cellcolor[HTML]{1F77B4}10 & \cellcolor[HTML]{F2F2F2}{\scriptsize--} & \cellcolor[HTML]{1F77B4}10 & \cellcolor[HTML]{4B92C3}9 & \cellcolor[HTML]{1F77B4}10 & \spk{10}{10}{9}{10} \\
  \quad 3 & \cellcolor[HTML]{FFFFFF}5 & \cellcolor[HTML]{E8720C}0 & \cellcolor[HTML]{F2F2F2}{\scriptsize--} & \cellcolor[HTML]{E8720C}0 & \cellcolor[HTML]{FAE2CE}4 & \spk{5}{0}{0}{4} \\
  \quad 4 & \cellcolor[HTML]{EC8E3C}1 & \cellcolor[HTML]{E8720C}0 & \cellcolor[HTML]{F2F2F2}{\scriptsize--} & \cellcolor[HTML]{78ADD2}8 & \cellcolor[HTML]{78ADD2}8 & \spk{1}{0}{8}{8} \\
  \quad 5 & \cellcolor[HTML]{78ADD2}8 & \cellcolor[HTML]{F2F2F2}{\scriptsize--} & \cellcolor[HTML]{E8720C}0 & \cellcolor[HTML]{FAE2CE}4 & \cellcolor[HTML]{1F77B4}10 & \spk{8}{0}{4}{10} \\
  \quad 6 & \cellcolor[HTML]{FAE2CE}4 & \cellcolor[HTML]{4B92C3}9 & \cellcolor[HTML]{F2F2F2}{\scriptsize--} & \cellcolor[HTML]{4B92C3}9 & \cellcolor[HTML]{78ADD2}8 & \spk{4}{9}{9}{8} \\
  \quad\textit{total} & \textbf{38} & \textbf{9} & \textbf{20} & \textbf{40} & \textbf{50} & \\[3pt]
\multicolumn{7}{@{}l}{\textit{regex}} \\[-1pt]
  \quad 1 & \cellcolor[HTML]{E8720C}0 & \cellcolor[HTML]{F2F2F2}{\scriptsize--} & \cellcolor[HTML]{E8720C}0 & \cellcolor[HTML]{E8720C}0 & \cellcolor[HTML]{F5C69D}3 & \spk{0}{0}{0}{3} \\
  \quad 2 & \cellcolor[HTML]{F5C69D}3 & \cellcolor[HTML]{FFFFFF}5 & \cellcolor[HTML]{F2F2F2}{\scriptsize--} & \cellcolor[HTML]{E8720C}0 & \cellcolor[HTML]{F1AA6D}2 & \spk{3}{5}{0}{2} \\
  \quad 3 & \cellcolor[HTML]{F5C69D}3 & \cellcolor[HTML]{E8720C}0 & \cellcolor[HTML]{F2F2F2}{\scriptsize--} & \cellcolor[HTML]{E8720C}0 & \cellcolor[HTML]{EC8E3C}1 & \spk{3}{0}{0}{1} \\
  \quad 4 & \cellcolor[HTML]{FAE2CE}4 & \cellcolor[HTML]{F2F2F2}{\scriptsize--} & \cellcolor[HTML]{F1AA6D}2 & \cellcolor[HTML]{E8720C}0 & \cellcolor[HTML]{F1AA6D}2 & \spk{4}{2}{0}{2} \\
  \quad 5 & \cellcolor[HTML]{E8720C}0 & \cellcolor[HTML]{E8720C}0 & \cellcolor[HTML]{F2F2F2}{\scriptsize--} & \cellcolor[HTML]{E8720C}0 & \cellcolor[HTML]{F5C69D}3 & \spk{0}{0}{0}{3} \\
  \quad 6 & \cellcolor[HTML]{E8720C}0 & \cellcolor[HTML]{EC8E3C}1 & \cellcolor[HTML]{F2F2F2}{\scriptsize--} & \cellcolor[HTML]{EC8E3C}1 & \cellcolor[HTML]{FAE2CE}4 & \spk{0}{1}{1}{4} \\
  \quad\textit{total} & \textbf{10} & \textbf{6} & \textbf{2} & \textbf{1} & \textbf{15} & \\[3pt]
\multicolumn{7}{@{}l}{\textit{topo-sort}} \\[-1pt]
  \quad 1 & \cellcolor[HTML]{F1AA6D}2 & \cellcolor[HTML]{EC8E3C}1 & \cellcolor[HTML]{F2F2F2}{\scriptsize--} & \cellcolor[HTML]{E8720C}0 & \cellcolor[HTML]{F1AA6D}2 & \spk{2}{1}{0}{2} \\
  \quad 2 & \cellcolor[HTML]{F5C69D}3 & \cellcolor[HTML]{F2F2F2}{\scriptsize--} & \cellcolor[HTML]{FAE2CE}4 & \cellcolor[HTML]{A5C8E1}7 & \cellcolor[HTML]{FAE2CE}4 & \spk{3}{4}{7}{4} \\
  \quad 3 & \cellcolor[HTML]{F5C69D}3 & \cellcolor[HTML]{FFFFFF}5 & \cellcolor[HTML]{F2F2F2}{\scriptsize--} & \cellcolor[HTML]{F1AA6D}2 & \cellcolor[HTML]{D2E3F0}6 & \spk{3}{5}{2}{6} \\
  \quad 4 & \cellcolor[HTML]{1F77B4}10 & \cellcolor[HTML]{F2F2F2}{\scriptsize--} & \cellcolor[HTML]{1F77B4}10 & \cellcolor[HTML]{1F77B4}10 & \cellcolor[HTML]{1F77B4}10 & \spk{10}{10}{10}{10} \\
  \quad 5 & \cellcolor[HTML]{E8720C}0 & \cellcolor[HTML]{FFFFFF}5 & \cellcolor[HTML]{F2F2F2}{\scriptsize--} & \cellcolor[HTML]{F5C69D}3 & \cellcolor[HTML]{FFFFFF}5 & \spk{0}{5}{3}{5} \\
  \quad 6 & \cellcolor[HTML]{F1AA6D}2 & \cellcolor[HTML]{1F77B4}10 & \cellcolor[HTML]{F2F2F2}{\scriptsize--} & \cellcolor[HTML]{1F77B4}10 & \cellcolor[HTML]{1F77B4}10 & \spk{2}{10}{10}{10} \\
  \quad\textit{total} & \textbf{20} & \textbf{21} & \textbf{14} & \textbf{32} & \textbf{37} & \\[3pt]
\bottomrule
\end{tabular}
\end{minipage}

\caption{Per-chaff success counts out of 10 runs (qwen2.5-coder:14b).
  Cell shading: continuous orange--white--blue gradient
  (orange $= 0/10$ runs, white $= 5/10$, blue $= 10/10$).
  The rightmost column is a sparkline~\cite{tufte2006sparklines} showing the success
  profile across the four underlying modes \lstinline|untyped|,
  \lstinline|min|, \lstinline|proximate|, \lstinline|all|;
  an upward slope means more verbose error messages helped on that chaff.
  Row totals across all 600 trials:
  \lstinline|untyped| 201,
  \lstinline|min-no-loc| 97,
  \lstinline|min-with-loc| 153,
  \lstinline|proximate| 287,
  \lstinline|all| 317.}
\label{tab:chaff-pivot}
\end{figure}

This statistical trend, however, masks real variation under the
hood. The variation shows that we cannot uniformly assume that one
mode of reporting will always be better than the other.

We summarize this using the table in \cref{tab:chaff-pivot}. In this table, each row is
a chaff. We see how it performed across ten runs; the number indicates
how many times the program was successfully fixed. We see that:
\begin{itemize}

\item Some chaffs are extremely easy across all modes (e.g.,
  \lstinline|rd-parser-1|, \lstinline|rd-parser-2|,
  \lstinline|topo-sort-4|, \lstinline|poly-arith-2|).

\item Some chaffs are extremely hard across all modes (e.g.,
  \lstinline|avl-tree-6|, \lstinline|functional-queue-4|).

\item \lstinline|regex| is hard across the board.

\item A few chaffs exhibit striking inversions of the above trend:
  e.g., \lstinline|avl-tree-1|, \lstinline|avl-tree-3|,
  \lstinline|poly-arith-1|, \lstinline|leftist-heap-4|,
  \lstinline|huffman-6|, \lstinline|dijkstra-2|.

\end{itemize}
Space precludes a detailed analysis of all these programs.

A natural question is whether the error-fixing performance correlates
with the distance of the \lstinline|proximate| error report from the
true error location. Because we did not set out to study this question
initially, we refrain from computing it here (to avoid
$p$-hacking). However, for the curious reader, we do note in passing
that we do not see any particular connection: the two don't seem
related.

\subsection{\RQref{types}: The Role of the Type System}
\label{sec:rq-types}

\RQref{types} asks whether having a type system in the feedback loop
helps the agent converge on a correct fix, compared to relying on test
output alone. The typed modes---especially \lstinline|proximate| and
\lstinline|all|---consistently outperform \lstinline|untyped|.

Causal attribution requires care, however. The \lstinline|untyped|
condition differs from the typed conditions in two ways simultaneously:
it removes type-error feedback \emph{and} it changes the nature of the
feedback signal from a type-error message to raw test output. An agent
that struggles in \lstinline|untyped| mode may be reacting to the
absence of the structured type-error signal, not the absence of the
type system per se. Disentangling these two effects would require a
condition that provides test-only feedback \emph{with} type
information somehow embedded---a design we did not pursue here.

With that caveat, we can speculate on a possible explanation. The type
error message is a \emph{causal} signal---it names the conflicting
types and, in richer modes, the expressions that produced them---while
test failure is a \emph{symptomatic} signal: some output was wrong,
but the agent must reason backwards to identify the source. A single
type error also essentially summarizes what is learned from several
test failures.  Given that agents in all modes rarely needed more than
one turn to succeed (\cref{sec:rq-verb}), the advantage of typed
feedback does not come from tighter iterative convergence; it seems to
come from enabling a better first hypothesis. The type error message
appears to tell the agent \emph{what} went wrong and (in typed modes
with location information) \emph{where}; the test output primarily
tells it \emph{that} something went wrong.

\subsection{\RQref{folk}: Type Correctness and Semantic Correctness}
\label{sec:rq-folk}

\RQref{folk} asks how often an agent that fixes the type error also
passes the semantic tests. Across all typed modes and all 10 runs,
there were 872 trials in which the agent's final submission was
type-correct (i.e., the oracle's type-check step passed). Of those,
854 also passed all semantic tests---a rate of 97.9\,\%.

This is strikingly high. But the caveat is equally important: our
chaffs are deliberately simple single-error programs, each with one
and only one deliberate type mistake. An agent that finds and reverses
that one change is very likely to restore the original correct
behavior, precisely because nothing else was broken to begin with. The
97.9\,\% figure may say more about our chaff design than about any
fundamental property of typed languages in general. In programs with
multiple interacting errors, type-correctness and semantic correctness
could diverge significantly.

Given that caveat, the result does provide preliminary support for the
folk belief: in a corpus of deliberately simple, single-error
programs, getting a type-checking tool to accept the code is an almost
perfectly reliable signal that the fix is also semantically
correct. This could make type-directed feedback especially valuable
for AI agents, especially since it is likely to be much quicker than
running a large test suite.

\subsection{Using State-of-the-Art Models}
\label{sec:found-mod}

It is natural to wonder if the results would be different if we had
used state-of-the-art models instead. We fully expect they would be:
the programs were generated by such models, so it would not
be surprising if the models could also debug such programs. Also, at
this point, small programs for well-known tasks, of the kind in
\cref{tab:programs}, should
be very well represented in the training sets.
The experiments described below cost a total of USD 171 (a fair bit
lower than that suggested in \cref{sec:agents} thanks to using a
cheaper model, adding thresholds, using aider instead of a
hand-written agentic loop, and several other changes).

To confirm our belief, we configured aider to run against {\haiku}, a
fairly fast and lower-cost alternative to the leading model from
Anthropic at the time of this writing (4.8). We ran the same workload
against {\haiku} as we had against {\qwen}. \footnote{We used a 300
  second timeout instead of 600s due to the much faster speed of
  interaction. Either way, it is not really meaningful to
  compare across these two very different setups.}
Unsurprisingly, we saw very high success rates
(\cref{tab:outcomes-haiku}).

\begin{table}[t]
  \centering
  \caption{Outcome distribution for {\haiku}. The columns are the same as in \cref{tab:outcomes-qwen}.}
  \label{tab:outcomes-haiku}
  \begin{tabular}{@{}lr rrr rrr@{}}
    \toprule
    & & \multicolumn{3}{c}{Timed out} & \multicolumn{3}{c}{Halted} \\
    \cmidrule(lr){3-5}\cmidrule(lr){6-8}
    Mode & Success & \makecell{tests\\failed} & \makecell{type\\error} & \makecell{no\\change} & \makecell{no\\edit} & stopped & \makecell{reflection\\cap} \\
    \midrule
    \lstinline|untyped| & 90.4/95.0 & 0.0/0.0 &  \multicolumn{1}{c}{---} & 0.0/0.0 & 4.6/1.2 & 0.4/0.9 & 4.6/1.9 \\
    \lstinline|min-...| & 90.0/96.2 & 1.4/0.3 & 0.0/0.3 & 0.0/0.0 & 1.8/1.9 & 0.4/0.0 & 6.4/0.9 \\
    \lstinline|proximate| & 87.9/99.7 & 1.4/0.0 & 0.0/0.0 & 0.0/0.0 & 4.3/0.0 & 0.0/0.0 & 6.4/0.0 \\
    \lstinline|all| & 92.5/98.4 & 0.7/0.6 & 0.0/0.0 & 0.0/0.0 & 2.9/0.6 & 0.4/0.0 & 3.6/0.0 \\
    \bottomrule
  \end{tabular}
\end{table}

\paragraph{Program Obfuscation}

The wheat versions of the programs have meaningful names, and the
error-injected chaffs retain the same names. These are then seen by
the agent---e.g., ``\lstinline|avl_inorder|''---as shown in
\cref{sec:modes}. There is some question of whether this is ``fair''
or not. On the one hand, these are well-known problems so the models
have probably seen numerous solutions during training. On the other
hand, this was not sufficient to obtain immediate fixes for
{\qwen}. Furthermore, language models benefit in part from the names
used in programs to infer intent.

Nevertheless, given the ease with which {\haiku} solved these
problems, we constructed ``obfuscated'' versions
of the chaffs by $\alpha$-renaming all the names---of types, of
fields, of variables---with meaningless two-letter names.\footnote{We
  note in passing the difficulty of performing this obfuscation
  well. Even if the file contents are scrubbed, there are other
  sources that contain information, such as filenames and
  pathnames. These can then subtly be leaked back to the model: e.g.,
  in the type or test failure error report that is sent to the model
  to revise the program. Through careful manual inspection and a
  series of searches we found and were able to eliminate multiple of
  these, all of which had been missed by AI agents
  (\cref{sec:ai-use}). This points to a danger of excessive automation
  and the potential for subtly incorrect or even wrong results.}
Thus, for instance, the AVL tree definition in \cref{sec:shplait}
turned into\footnote{The renaming was done by a program that then
  pretty-printed the results, so the renamed versions are
  syntactically slightly different from the originals.}
\begin{lstlisting}
type Aa (? a)
| ah ()
| ap (bd :: Int, df :: ? a, az :: Int, bh :: Aa (? a), cg :: Aa (? a))
\end{lstlisting}
and an AVL height function became
\begin{lstlisting}
fun aj (cu):
  match cu
  | ah (): 0
  | ap (ab, ag, ay, bf, ae): bf
\end{lstlisting}

At some fundamental level, our obfuscation did not entirely succeed.
Looking at the model's printed output, we can see clear evidence that
the model is able to reconstruct the problem: e.g.,
\begin{verbatim}
 The code appears to be implementing Dijkstra's algorithm                     
\end{verbatim}
or
\begin{verbatim}
 ax is Kleene star (zero or more)                                             
\end{verbatim}
Indeed, a simple search is revealing. Each program contributes 240
trials. Searching in the experimental output log (and spot-checking
several of these to confirm that they are meaningful and relevant to this
discussion) reveals 53 occurrences of
``AVL'', 18 of ``Dijkstra'', 18 of ``Huffman'', 25 of ``leftist'', 24
of ``topological'', and 204 [sic] of ``Kleene''. Some of these are
duplicates per run, so these numbers should not be
over-interpreted. Our point is only that the non-trivial presence of
these tell-tale words is a clear indication that the models are often
able to reconstruct what program they are processing---and use that
knowledge to fix bugs---despite the lack of any useful names at
all.\footnote{A machine learning colleague of the first author
  suggested that large commercial models may be trained to work with
  obfuscated names, to make them more robust.}
Nevertheless, making this change did have a non-trivial effect on
success: see \cref{tab:outcomes-haiku-obf}.

\begin{table}[t]
  \centering
  \caption{Outcome distribution for {\haiku} obfuscated. The columns are the same as in \cref{tab:outcomes-qwen}.}
  \label{tab:outcomes-haiku-obf}
  \begin{tabular}{@{}lr rrr rrr@{}}
    \toprule
    & & \multicolumn{3}{c}{Timed out} & \multicolumn{3}{c}{Halted} \\
    \cmidrule(lr){3-5}\cmidrule(lr){6-8}
    Mode & Success & \makecell{tests\\failed} & \makecell{type\\error} & \makecell{no\\change} & \makecell{no\\edit} & stopped & \makecell{reflection\\cap} \\
    \midrule
    \lstinline|untyped| & 59.3/65.9 & 5.0/3.4 &  \multicolumn{1}{c}{---} & 0.0/0.0 & 3.2/3.4 & 2.5/1.2 & 28.2/25.9 \\
    \lstinline|min-...| & 43.6/63.1 & 1.4/2.8 & 3.9/0.9 & 0.0/0.0 & 6.4/4.7 & 3.6/1.2 & 40.7/27.2 \\
    \lstinline|proximate| & 60.4/70.6 & 1.8/2.5 & 2.9/0.0 & 0.0/0.0 & 3.9/4.4 & 2.5/0.6 & 28.6/21.9 \\
    \lstinline|all| & 64.3/74.4 & 6.1/3.4 & 2.5/0.3 & 0.0/0.0 & 2.5/0.9 & 2.9/1.9 & 21.1/18.4 \\
    \bottomrule
  \end{tabular}
\end{table}

\paragraph{Statistical Analysis}

Because each {\haiku} trial is an independent API call with no shared
per-run hardware, we do not block by run; we instead test the
predicted ordering with the Jonckheere-Terpstra test for ordered
alternatives across independent
samples~\cite{jonckheere1954distribution, terpstra1952asymptotic}, the
unblocked counterpart of Page's test. We run
the test on each chaff bin separately for both {\haiku} conditions.
Three of the four tests are significant at $p < 0.05$.
Without obfuscation: for min-no-loc, $z = 0.54$, $p = 0.30$,
one-sided (\emph{not} significant); for min-with-loc, $z = 3.42$,
$p = 3.1 \times 10^{-4}$, one-sided. With obfuscation: for min-no-loc,
$z = 2.40$, $p = 8.2 \times 10^{-3}$, one-sided; for min-with-loc,
$z = 2.82$, $p = 2.4 \times 10^{-3}$, one-sided. Nevertheless, the
unobfuscated code has an extremely high baseline, and the obfuscated
case has a non-monotonic dip in both bins. We therefore present these
solely as points of potential interest for future exploration.

\section{Execution Statistics}
\label{sec:exec-stats}

\begin{table}[t]
  \centering
  \caption{Per-trial token usage, elapsed time, and turns, by model and verbosity mode. M is mean and Mdn is median.}
  \label{tab:tokens-and-time}
  \begin{tabular}{lrrrrrrrrr}
    \toprule
     &  & \multicolumn{2}{c}{input tokens}  & \multicolumn{2}{c}{output tokens}  & \multicolumn{2}{c}{elapsed (s)}  & \multicolumn{2}{c}{turns} \\
    \cmidrule(lr){3-4} \cmidrule(lr){5-6} \cmidrule(lr){7-8} \cmidrule(lr){9-10}
    Mode & \#trials & M & Mdn & M & Mdn & M & Mdn & M & Mdn \\
    \midrule
    \multicolumn{10}{l}{\textbf{\texttt{claude-haiku-4-5-20251001}}} \\
    \lstinline|untyped| & 600 & 14{,}674 & 13{,}000 & 543 & 422 & 57.9 & 45.0 & 1.2 & 1.0 \\
    \lstinline|min-no-loc| & 280 & 8{,}954 & 7{,}900 & 585 & 464 & 74.8 & 64.0 & 1.4 & 1.0 \\
    \lstinline|min-with-loc| & 320 & 8{,}651 & 7{,}900 & 467 & 419 & 73.1 & 62.5 & 1.1 & 1.0 \\
    \lstinline|proximate| & 600 & 8{,}580 & 7{,}900 & 454 & 397 & 72.8 & 61.0 & 1.2 & 1.0 \\
    \lstinline|all| & 600 & 8{,}737 & 8{,}100 & 425 & 369 & 72.9 & 59.0 & 1.1 & 1.0 \\
    \addlinespace
    \textit{all} & 2400 & 10{,}196 & 8{,}100 & 486 & 406 & 69.4 & 56.0 & 1.2 & 1.0 \\
    \midrule
    \multicolumn{10}{l}{\textbf{\texttt{claude-haiku-4-5-20251001-obf}}} \\
    \lstinline|untyped| & 600 & 35{,}618 & 16{,}000 & 1{,}200 & 830 & 92.3 & 69.0 & 2.1 & 2.0 \\
    \lstinline|min-no-loc| & 280 & 19{,}890 & 8{,}100 & 1{,}367 & 1{,}128 & 115.0 & 89.5 & 2.5 & 3.0 \\
    \lstinline|min-with-loc| & 320 & 18{,}993 & 8{,}100 & 1{,}142 & 801 & 92.1 & 74.5 & 2.1 & 2.0 \\
    \lstinline|proximate| & 600 & 17{,}391 & 8{,}100 & 991 & 688 & 102.4 & 82.0 & 2.0 & 1.0 \\
    \lstinline|all| & 600 & 18{,}203 & 8{,}300 & 877 & 562 & 106.1 & 79.0 & 1.9 & 1.0 \\
    \addlinespace
    \textit{all} & 2400 & 22{,}656 & 11{,}000 & 1{,}078 & 762 & 100.9 & 77.0 & 2.1 & 2.0 \\
    \midrule
    \multicolumn{10}{l}{\textbf{\texttt{qwen2.5-coder:14b}}} \\
    \lstinline|untyped| & 600 & 11{,}384 & 11{,}000 & 392 & 310 & 209.4 & 144.0 & 1.9 & 1.0 \\
    \lstinline|min-no-loc| & 280 & 7{,}171 & 7{,}100 & 295 & 230 & 161.2 & 124.0 & 2.4 & 2.0 \\
    \lstinline|min-with-loc| & 320 & 7{,}124 & 7{,}100 & 265 & 238 & 127.6 & 92.5 & 1.8 & 1.0 \\
    \lstinline|proximate| & 600 & 7{,}177 & 7{,}100 & 252 & 237 & 133.1 & 96.5 & 1.8 & 1.0 \\
    \lstinline|all| & 600 & 7{,}380 & 7{,}300 & 248 & 224 & 134.4 & 102.0 & 1.8 & 1.0 \\
    \addlinespace
    \textit{all} & 2400 & 8{,}272 & 7{,}300 & 293 & 246 & 155.0 & 110.0 & 1.9 & 1.0 \\
    \bottomrule
  \end{tabular}
\end{table}

Finally, in \cref{tab:tokens-and-time}, we present the number of
tokens, and time, spent on each trial.

We can see in this the larger numbers of output tokens generated by
{\haiku} (reflecting the longer analysis it provides), and how the
obfuscated versions take many more tokens and more turns. The
\lstinline|untyped| modes need more tokens because the entire test
failure log is sent to the model. The numbers show per \emph{trial}
usage, which is naturally higher when there are more
turns. Furthermore, each turn includes the prior conversation history,
so multiple turns lead to even more token growth.

The timing differences between {\haiku} and {\qwen} likely reflect
compute stack differences. Those between the two {\haiku} conditions
presumably reflect the cost of obfuscation.

\section{Threats to Validity}
\label{sec:threats}

This work must be considered very preliminary. We primarily intend for
it to spark discussion and inspire future work, not to provide
definitive answers. Below, we list concrete threats in more detail.

\begin{description}

  \item[Internal validity] At least in the \lstinline|untyped|
    condition, we have made two changes (no type system combined with
    a different kind of feedback). Therefore, we cannot cleanly
    attribute the results to the mode.

  \item[External validity] This is obviously the biggest threat. We
    have used only one language, small programs, a single model,
    single-error chaffs, ten runs, and so on. The results can in no
    way be considered generalizable.

  \item[Construct validity] We define success by passing the test
    suite. While this is not the same as careful, manual inspection
    for correctness, we believe this is a common measure in many
    software engineering settings. We took care both to keep the tests
    away from the model, and to ensure that several test failures had
    to be overcome before declaring success. Nevertheless, there may
    be subtle weaknesses in the tests that make this less meaningful
    as a metric.

  \item[Reliability] We have taken many steps, especially using an
    open-weight model, to enable reproducibility. Nevertheless, given
    the stochastic nature of language models, and differences in
    hardware (which can impact timeouts), we cannot be sure that the
    results will replicate.

\end{description}

\section{Background and Related Work}
\label{sec:rel-work}

\paragraph{Type inference error reporting.}
There is a long history of work on trying to improve error reporting in
ML-style type inference systems, starting with Wand~\cite{Wand1986}.
For instance, Lerner et al.~\cite{lerner2007searching} search the space
of edits to produce more helpful messages; Heeren et
al.~\cite{heeren2003scripting} let library authors script the type-inference and
reporting process (the Helium compiler); Haack and Wells~\cite{haack2004slicing}
compute type-error \emph{slices} that report every program point contributing to
a conflict; and Wu et al.~\cite{wu2017userfriendly} learn user-friendly messages.
Hage~\cite{hage2020solved} surveys the open
problems in this space.

\paragraph{Type-error localization and blame.}
Closest to our blame-distance analysis, Seidel et al.~\cite{seidel2017blame}
learn a data-driven model that predicts which sub-expression to \emph{blame} for
a (novice) type error, outperforming the compiler's own localization. Where they
measure blame accuracy for human learners, we measure how far an LLM agent's
repair lands from the line at which the fault was actually injected. Geng et
al.~\cite{geng2022novice} carry this diagnosis task to (natural-)language
models---a step toward our agentic LLM setting.

\paragraph{Do error messages help?}
For human programmers the evidence is mixed. Marceau et al.~\cite{mfk:measur-effect-error-msg-novice-sigcse}
study human responses to error messages to identify problems with
error reporting.
Becker~\cite{becker2016effective}
reports that enhanced compiler error messages help novices; Denny et
al.~\cite{denny2014enhancing} find enhancement ineffectual in a controlled
study; and the working-group survey of Becker et al.~\cite{becker2019unhelpful}
concludes that the picture remains unsettled. We pose the analogous, controlled
question for an LLM agent rather than a student.

\paragraph{Comprehension of degraded programs.}
Our obfuscated condition connects to work on how degrading a program affects
comprehension, reminiscent of the chess-perception paradigm in
which an expert's recall advantage for real boards collapses for randomly
arranged ones~\cite{degroot1965thought,chase1973perception}. \emph{Line scrambling}:
Shneiderman~\cite{shneiderman1976exploratory} and McKeithen et
al.~\cite{mckeithen1981knowledge} randomly reorder statements; the expert
advantage shrinks but---unlike the strict chess result---is not
eliminated. \emph{Plan/discourse violation}: Soloway and
Ehrlich~\cite{soloway1984empirical} contrast programs that obey programming plans
and conventions with ones that compute the same result while violating them
(their canonical example is a variable named \texttt{max} that holds the
minimum); experts far outperform novices on the conventional version but perform
essentially the same as novices on the violating one. \emph{Identifier
degradation}: Teasley~\cite{teasley1994effects}---the only study to vary
identifier names alone, the program otherwise intact, building on Pennington's
model of program comprehension~\cite{pennington1987stimulus}---finds a main
effect of expertise but a more muted picture, with no clean expert--novice
crossover. Our obfuscation most resembles the last of these. Of
course, all the other studies were conducted on expert programmers,
not on language models.

\paragraph{Automated program repair.}
Automatically repairing buggy programs has a long history.  Shapiro's
seminal dissertation~\cite{shapiro1983algorithmic} interactively
localized faults by querying the programmer about intermediate
values (but not, by itself, proposing a fix).
Modern automated program repair
synthesizes a patch that makes a failing test suite pass: the
search-based formulation of GenProg~\cite{legoues2012genprog} is an
early work, followed by several other methods~\cite{legoues2019survey}.
Pre-trained code models have substantially recast the
pipeline, with Xia et~al.~\cite{xia2023llmrepair}
providing a large-scale evaluation.  Our work sits inside
this latter setting but asks an orthogonal question: rather than
which prompting or fine-tuning strategy best improves repair, we hold
the agent fixed and vary the typed context.

\section{Discussion}
\label{sec:discussion}

Programming languages, from their inception, have been designed for
humans. Grace Hopper's FLOW-MATIC (which inspired COBOL) was marketed on
``the use of English words''~\cite{flowmatic}. As early as 1976 (and
probably earlier), researchers were focused on how error messages can
help humans~\cite{Horning1976}. Decades of work have gone into the
\emph{human} factors of programming, including the presentation of
errors.

Now we have an entirely new target audience: AI agents. They have
different characteristics, especially when it comes to reading
``prose''. It is therefore important to ask whether they can be better
served by different kinds of output. This paper represents a very
preliminary investigation into this matter.

We have constructed an experiment where a language provides
progressively greater type error information (starting with none at
all, and only dynamic errors). We use modest programs, and a
correspondingly modest model (such that the model has moderate success
across the modes: neither total failure nor complete success). With
that, we find that richer type information generally leads to better automated
program repair success. While these problems are too simplistic for a
state-of-the-art model, we see similar behavior when we obfuscate the
programs on that same model.

Along the way, we have also made various observations that we believe
might be worth investigating further:
\begin{itemize}

\item We see some evidence that types help over just dynamic (safety
  and) test failures.

\item We see some evidence for the folk claim that ``when the types
  check, the program is correct''.

\item We see clear evidence that state-of-the-art models are able to
  infer (very well-known) algorithms even when the names have been
  completely obfuscated.

\item Untyped programs use many more tokens (which is unsurprising in
  our setup: we have not performed ablations to see how many tests are
  sufficient to achieve effective debugging), but for {\haiku}
  seem to take less time per trial.

\end{itemize}

We reiterate that this is meant to be a preliminary study meant to
spur more research in this area. At the very least, we believe that
there is real evidence for having different output modes for humans
and machines (specifically, agentic AI systems).

\section{AI Use}
\label{sec:ai-use}

We now disclose the uses of AI.
All the work below was done with Anthropic Claude
4.6 (Sonnet) and 4.7 (Opus).  One author (SK) was responsible for all
this use.

\begin{description}

\item[problem generation] The author did several rounds of iteration
  with Claude to generalize his ideas for problems (such as
  \lstinline|avl-tree|) as well as categories of problems (such as
  graph algorithms). He eventually chose the final set of problems.

\item[wheat creation] For ``frontier'' models, generating solutions
  for these problems is quite straightforward. The author guided
  Claude to work correctly with Shplait. He also made decisions such
  as having type annotations in the datatype declarations but none in
  the rest of the program. He then checked that the programs looked
  credible.

  \item[test-suite generation] The author had Claude generate test
    suites, and examined them for enough detail, forcing the model to
    iterate until they had reasonable coverage.

  \item[chaff creation] The author worked iteratively with Claude to
    generate chaffs. He guided it to think about mistakes versus
    misconceptions for each problem, forced it to consider the
    symmetric cases, and chose the final set of chaffs. He then guided
    it to generate the corresponding code as a modification of the
    wheat, confirming that it met the criteria desired for the
    experiment.

  \item[test-suite hardening] The author employed Claude iteratively
    to made sure the test suites were hardened to have at least five
    failures for each chaff.

  \item[experimental harness creation] The author had Claude provide
    guidance on model selection and open-source agent software. He
    then had it write the code to invoke these in conjunction.

  \item[deployment on GPU cluster] After testing the code on his
    personal laptop over multiple days and nights, the author used
    Claude to set up the task on a GPU cluster. In particular, Claude
    was useful for performing tasks like machine selection (based on
    their architecture) and debugging speed and completion issues
    (such as the configuration halting prematurely or running on a CPU
    instead of a GPU).

  \item[generation of summary statistics] The author used Claude to
    write the straightforward but painstaking scripts to generate data
    summaries from the execution logs.

  \item[generation of charts] The author used Claude to generate
    scripts that would turn the data summaries into LaTeX tables,
    delegating to it to handle various LaTeX formatting issues.

\end{description}
At all points the author using Claude has examined the output and done
his best to validate it, exactly the same as if the above tasks had
been done by a PhD student.

\paragraph{Artifact}

The paper's artifact is available from Zenodo at
\url{https://doi.org/10.5281/zenodo.20481462} (DOI:
\texttt{10.5281/zenodo.20481462}).

\acks

We are deeply grateful to: Will Crichton for a careful reading, Kathi
Fisler for pointing us to related work, Stephen Bach for a useful
discussion, and Yaron Minsky for a tweet that pushed us to speed up
our work on this project.
This work is partially funded by US NSF grant CCF-2227863.

\newpage

\bibliography{related-work,sk}

\end{document}